\documentstyle[12pt,aasms4,psfig]{article}

\newcommand{\gsim}{\mbox{\hspace{.2em}\raisebox{.5ex}{$>$}\hspace{-.8em}\raisebox{-.5ex}{$\sim$}\hspace{.2em}}}  % \newcommand{\gsim}{\stackrel{>}{\sim}}
\newcommand{\ssst}{\scriptscriptstyle}
\newcommand{\E}[1]{\times 10^{#1}}
\newcommand{\RA}[3]{\mbox{R.A.}={#1}^{{\rm h}}{#2}^{{\rm m}}{#3}^{{\rm s}}}
\newcommand{\decl}[3]{\mbox{decl.}={#1}^{\circ}{#2}'{#3}''}

\newcommand{\s}{\,{\rm s}} 	
\newcommand{\yr}{\,{\rm yr}}	
\newcommand{\cm}{\,{\rm cm}}	
\newcommand{\cmps}{\cm\s^{-1}}
\newcommand{\parsec}{\,{\rm pc}}\newcommand{\kpc}{\,{\rm kpc}}
\newcommand{\ergs}{\,{\rm ergs}}	
	\newcommand{\Msolar}{M_{\odot}}
	\newcommand{\keV}{\,{\rm keV}}
\newcommand{\G}{\,{\rm G}}

\newcommand{\nel}{n_{e}}	\newcommand{\NH}{N_{\ssst H}}
\newcommand{\nHc}{n_{{\ssst H},c}} \newcommand{\nHicm}{n_{{\ssst H},{\ssst ICM}}}
\newcommand{\pcl}{p_{c}}	\newcommand{\picm}{p_{\ssst ICM}}
\newcommand{\Ts}{T_{s}}		\newcommand{\vs}{v_{s}}
 	\newcommand{\mH}{m_{\ssst H}}

\newcommand{\mE}{\langle E_{\ssst B-V}\rangle}
\newcommand{\xray}{X-ray}
\newcommand{\Einstein}{{\em Einstein}}
\newcommand{\ROSAT}{{\em ROSAT}} \newcommand{\ASCA}{{\em ASCA}}
\newcommand{\AXAF}{{\em AXAF}}	\newcommand{\XMM}{{\em XMM}}

\begin{document}
\lefthead{Chen et al.}
\righthead{\xray\ Emission of SNR 3C~397}

\title{AN ANALYSIS OF THE X-RAY EMISSION FROM THE SUPERNOVA REMNANT 3C~397}

\author{Yang Chen, Ming Sun, Zhen-Ru Wang}
%\altaffiltext{1}{Center of Astronomy and Astrophysics, CCAST(World Laboratory),
\affil{Department of Astronomy, Nanjing University, Nanjing 210093, P.R.China\\
Electronic mail: ygchen, quqiny, zrwang@nju.edu.cn}
\centerline{and}
\author{Q.~F.~Yin}
\affil{National Radio Astronomy Observatory\altaffilmark{1}, 520
Edgemont Road, Charlottesville, VA 22903\\Electronic mail: 
qyin@nrao.edu}
\altaffiltext{1}{The National Radio Astronomy Observatory is a
facility of the National Science Foundation operated under cooperative
agreement by Associated Universities, Inc.}

\vfil
%Text
\begin{abstract}

The \ASCA\ SIS and the \ROSAT\ PSPC spectral data of the SNR 3C~397
are analysed with a two-component non-equilibrium ionization model.
Besides, the \ASCA\ SIS0 and SIS1 spectra are also fitted simultaneously in an equilibrium case.
The resulting values of the hydrogen column density yield a distance of $\sim8\kpc$ to 3C397.
It is found that the hard \xray\ emission, containing S and Fe~K$\alpha$ lines,
arises primarily from the hot component,
while most of the soft emission, composed mainly of Mg, Si, Fe~L lines, and continuum,
is produced by the cool component.
The emission measures suggest that the remnant evolves in a cloudy medium
and imply that the supernova progenitor might not be a massive early-type star.
The cool component is approaching ionization equilibrium.
The ages estimated from the ionization parameters and dynamics are all much greater than the previous determination.
%We did not find powerlaw, bremsstrahlung, and BB components, as well as pulsed signals.
We restore the \xray\ maps using the \ASCA\ SIS data and compare
them with the \ROSAT\ HRI and the NRAO VLA Sky Survey (NVSS) 20 cm 
maps.
%The bipolar morphology may imply that the remnant encounters a denser medium in the west.
The morphology with two bright concentrations suggests a bipolar  remnant 
 encountering a denser medium in the west.


\end{abstract}
\section{Introduction}\label{sec:intro}
The bright radio and X-ray
extended source 3C~397 (also called G41.1$-$0.3 or HC~26) used to be
regarded as one of the youngest supernova remnants (SNRs).  It has
been observed by several telescopes for more than two
decades, and some preliminary understanding have been achieved.

The 20 cm observation of the Fleurs synthesis telescope first resolved
a shell structure with a small average diameter ($D\sim3'.6$), which
shows an apparent departure from spherical symmetry (Caswell et al.\
1982). The HII region G41.1$-$0.2 lies $6'$ to the west and in the
foreground (Caswell et al. 1975) and may extinguish the emission
from 3C~397 (Cersosimo \& Magnini 1990).  The HI absorption
measurements provide a distance estimate for 3C~397: $d\gsim7.5\kpc$
(Caswell et al.\ 1975), and the $\Sigma-D$ relationship suggests a
distance $\sim12$---$13\kpc$ (Caswell \& Lerche 1979; Milne 1979).  A
preliminary estimate of the remnant age is only $t\sim600\yr$ (Caswell et al.\
1982; Becker, Markert, \& Donahue 1985).  The spectral index of the
synchrotron emission was found to vary across the source without
evident correlation with the radio intensity features,  possibly due to
an inhomogeneous ambient medium (Anderson \& Rudnick
1993).

The \Einstein\ observation was reported by Becker et al.\ (1985).  Two
concentrations of soft \xray\ emission appear in the IPC image, and
Becker et al.\ suggested an association of some \xray\
emission with the central VLA radio component.  No stellar remnant was
indicated by the spectrum. A temperature $T<0.25\keV$ and a column
density of hydrogen $\NH>5\E{22}\cm^{-2}$ were estimated from the IPC
data while $T\sim1.0$---$6.0\keV$ and $\NH\le1.4\E{22}\cm^{-2}$ were
estimated from the MPC data.  However, Becker et al.\ did not think the
\Einstein\ data are of such high quality that these parameters are
determined reliably.

As yet the emission lines and element abundances were poorly known.
However, this situation could change and many of the basic physical properties
could be known more explicitly than before because of the recent \xray\
observations of 3C~397. In this paper, we first present the \xray\ spectral
analysis based on the archival \ASCA\ SIS data and \ROSAT\ PSPC data in
combination.  The \ROSAT\ HRI data, \ASCA\ SIS data, and NVSS 20 cm data
(Condon et al. 1998) are used for the investigation of the remnant morphology.
The data analysis is given in \S2, the physical features of the remnant is
discussed in \S3, and the results are concluded in \S4.

\section{Data Analysis}

The SNR 3C~397 was observed by the \ASCA\ satellite on April 8, 1995
with the SIS and GIS detectors.  The SIS data are in 1-CCD clocking
mode, and hence no residual dark distribution (RDD) correction is made.
After the standard screening process, the effective exposure times
were 34.1 kiloseconds for SIS0 and 33.7 kiloseconds for SIS1, and the
total events amount to 33943 and 27116 for SIS0 and SIS1,
respectively. 

The \ROSAT\ PSPC observation was made on October 28, 1992 with an
effective exposure time of $4165\s$.  The \ROSAT\ HRI observation
was made on October 21, 1992 and the effective exposure time was
$6153\s$.

The standard software FTOOL4.0, IRAF, and AIPS are used to process the 
\xray\ and NVSS data.

\subsection{Spectral Analysis}\label{sec:spec}
In view of the fact that the \ASCA\ SIS
has a better spectral resolution (2\% FWHM at 6 keV) than the GIS
(8\%), we use the SIS data in the spectral analysis.  The
spectrum of the remnant is extracted from the SIS data within a
circular region of radius $4'$  centered at
$\RA{19}{07}{33}$, $\decl{07}{08}{00}$
(J2000).  Background spectra are extracted from the archival data on
blank sky near 3C~397.  The distinct emission lines in the spectrum
are Mg~He$\alpha$ ($\sim 1.35\keV$), Si~He$\alpha$ ($\sim 1.85\keV$),
S~He$\alpha$ ($\sim 2.43\keV$), and the Fe~K$\alpha$ complex.  The Fe~L
complex also contributes considerably in the range about 0.65 ---
1.8$\keV$.  The centroid of the Fe~K$\alpha$ complex is 6.59 keV (with a 90\%
confidence range $\pm0.02\keV$, plus comparable errors caused by
calibration uncertainties, such as non-uniform charge transfer
inefficiency), which indicates the significance of  B-  through Li-like
iron ions. It is known that the distinct Fe~K$\alpha$ lines indicate
the existence of a high temperature ($\gsim1\keV$) plasma component,
and such a high temperature and a low centroid of the complex
indicate that this plasma component has not yet reached ionization
equilibrium (e.g. Borkowski \& Szymkowiak 1996).

The SIS0 spectrum is fitted with a two-component non-equilibrium ionization
(NEI) model using the spectral code SPEX (Kaastra
et al.\ 1996) in the energy range 0.5 --- 8.4 keV.  The model of
Morrison \& McCammon (1983) is used for the interstellar absorption
of the spectrum.  In order to avoid too many degrees of freedom in
fitting, we only allowed the abundances of those elements (Mg, Si, S,
\& Fe) showing distinct emission lines to vary during fitting.
Two cases, in which the elemental abundances of the two components
either vary independently or are coupled, are investigated.
The fitting results are listed in Table~1.  The reduced
$\chi^{2}\sim1.3$ in the independent case is better than the
$\chi^{2}\sim1.6$ in the coupled case.  Considering the \ROSAT\ PSPC has a
better response in the soft band ($<0.9\keV$) than the \ASCA, the combined
spectra of both the \ASCA\ SIS0 and the \ROSAT\ PSPC (0.2 --- 2 keV)
are fitted  with the NEI model again, and the results are very similar
to the pure SIS case (see Table~1).  No significant contributions of
power-law, bremsstrahlung, and blackbody components are found.

From Figure~1 one can find that the emission below about $2\keV$ 
arises predominantly from the cool component plasma, while the emission
above $2\keV$ arises mainly from the hot component.  The cool component
is responsible for the lines of Mg, Si, and Fe~L, while the hot one
responsible for the lines of S and Fe~K$\alpha$.  Noticeably, the
ionization parameter $\nel t\gsim1\E{12}\cm^{-3}\s$ of the cool component
(see Table~1) is so high that it indicates that the component is approaching
ionization equilibrium (e.g.\ Masai 1984).

Motivated by the possibility of equilibrium ionization (EI), we also
apply the XSPEC code (using the VMEKAL model therein) (Mewe et al.\
1995) to fit the SIS0 and SIS1 spectra in the range 0.5 ---
8.8 keV.  Besides the spectrum extracted from the circle
mentioned above, we also extract the spectra from the two circles
centered on positions A and B (labeled in Figure~2), each with a
radius of $1'.65$ covering an \xray\ concentration.
The fitting results are listed in Table 2.  Due to
the broad point spread function (PSF) of the \ASCA\ SIS and the small
separation of the two concentrations ($\sim 2'$), we must consider
the mutual contamination of the photons between regions A and B and
be careful about the results.  The signal-to-noise here is not very
good so that a sophisticated spatio-spectral analysis can hardly be
applied.  We use the \ROSAT\ PSPC data to check our results on
the interstellar absorption to A and B obtained from the \ASCA\ data.
The PSPC spectra are extracted from the same regions as the SIS spectra.
The statistical significance of the PSPC spectra is not very high, so we only
consider the energy band 0.8 --- 1.8 keV which is dominated by
the cool component (see Figure~1).  The temperature is fixed at 0.23
keV, the value obtained from \ASCA\ fitting.  The abundances of heavy
elements in the regions A and B are all fixed at the average values of
the SIS results.  Then we get $\NH(\rm A) =
2.83^{+0.13}_{-0.13}\times10^{22}$cm$^{-2}$ and $\NH(\rm B) =
3.06^{+0.15}_{-0.15}\times10^{22}$cm$^{-2}$, which are both very
similar to the SIS results. These results obtained from the PSPC and SIS
data favor the trend
$\NH({\rm A})<\NH({\rm B})$.  Furthermore, in spite of the
low statistical significance of the PSPC data, we still find sign
about the difference of Mg~He$\alpha$ emission between the spectra
of regions A and B.  More detailed spatio-spectral analysis of this
source could be an interesting subject for  future \AXAF\ and \XMM\
observations.

The low temperatures derived by the NEI (SPEX) and the EI (VMEKAL)
models are very similar, around $0.23\keV$, but the high temperatures
are quite different from one another.  This can be easily understood
due to the NEI nature of the hot component.  Notably, the low
temperatures in both models are in the temperature range determined ever
from the \Einstein\ IPC data and the high temperatures are in the range
determined from the MPC data (Becker et al.~1985). The abundances of Si in the
hot component are nearly zero either from the NEI model (except in the
coupled case) or from the EI model.  Both of the models yield high
values of the emission measure (EM) for the cool component compared
with those of the hot one.  Coincidentally, for the cool
component, the sum of the EMs of the regions A and B obtained from the EI
model are comparable with the EM obtained from the NEI model.

\subsection{Image Production}
Using the SIS data we make  maps in
the bands 0.5 --- 10 keV, 0.5 --- 2 keV, and 2 --- 10 keV.
The 0.5 --- 10 keV map with an overlay of the 20 cm radio emission contours
is shown in Figure~2.
The 2 --- 10 keV (hard) contours and the 20 cm radio emission contours
are superposed on the \ROSAT\ HRI colored image (Figure 3, Plate).
The reason here we select 2 keV to separate the soft map from the hard map is
that 2 keV is the approximate demarcation between the hot and cool components
in the spectral analysis (\S\ref{sec:spec}).  These SIS maps are produced
after the corrections for exposure and vignetting.
Here the ``Lucy-Richardson" method is applied for
40 iterations to deconvolve the \ASCA\ images with the PSF
(e.g. Jalota, Gotthelf, \& Zoonematkermani 1993).  
The 0.5 --- 10 keV image is very similar to the 0.5 --- 2
keV (soft) image because the 2 --- 10 keV (hard) emission from the
hot component is much weaker than the soft emission.  Both the 0.5 --- 10
keV and the soft images are similar to the \ROSAT\ PSPC image.
Here the 20 cm radio contour maps are made from the NVSS. The
resolution of the NVSS is 45 arcsec. (The rms brightness fluctuation of
the NVSS is $\sigma\approx 0.45$ mJy/beam$\approx 0.14$K [Stokes I].
The peak brightness is about 1.1 Jy/beam at
$\RA{19}{07}{27}.83$, $\decl{07}{08}{24}.31$ [J2000.0].)

\subsection{Search for the Pulsed Signal}
The GIS data are used in the temporal analysis owing to
their better time resolution.  After barycentering the photon times of
arrival, we extract the GIS2 (GIS3) light curves from the same
region as we extract the SIS spectra. Only
high bit-rate mode (resolution = 62.5 ms) data are used. Light curves
in the 0.5 --- 10 keV and 2 --- 10 keV bands are obtained. A $2^{20}$
point fast Fourier transform is applied to the light curves. No
significant pulsed signal is found in the range 0.125 --- 30 s.

\section{Discussion}

\subsection{Distance}
Both the NEI and EI fitting results give a
hydrogen column density $\NH$ around $\sim2.9\E{22}\cm^{-2}$ with a
range 2.6 --- 3.2$\E{22}\cm^{-2}$.  The extinction per unit distance
in the direction of 3C~397 can be estimated from the contour diagrams
given by Lucke (1978): $\mE/d\sim0.60\,{\rm mag}\kpc^{-1}$.  Using
the relation $\NH=5.9\E{21}\mE\cm^{-2}$ (Spitzer 1978), a distance
 $d\sim8.2\kpc$ (with a range 7.4 --- $9.0\kpc$) is obtained.  This
is in agreement with the limit $d\gsim7.5\kpc$ (Caswell et al.\ 1975).
Adopting an average angular radius of $2'$, the
radius of the remnant would be $r\sim4.7\parsec$.

\subsection{Emission Measure}

In a two-component model, the hot
component is usually ascribed to the surrounding medium swept up by the
blast wave and the cool component to the ejecta heated by the reverse
shock.  It seems not to be the case here, however, because the EM of the cool
component is much higher than that of the hot component.  A competing
factor may be the inhomogeneity of the surrounding medium.  The hot
component may correspond to the shocked low density intercloud matter
(ICM) and the cool component to the shocked dense cloud matter which
overwhelms the ICM in mass.  Similar cases were also encountered in other
remnants, such as the young SNR N132D (Favata et al.~1997).  For 3C~397,
in fact, inhomogeneities in the surrounding medium have been
invoked to account for the irregular distribution of the
radio spectral indices over the source, in view of the turbulence
stimulated by the onset of plasma instabilities when the ejecta
collide with the cloudlets (Anderson \& Rudnick 1993).

If the volume emission measure of the cool component is taken as $f\nel\nHc V\approx340\E{58}\cm^{-3}$, where $f$ is the filling factor of dense
cloudlets, then the hydrogen number density in the cloudlets is $\nHc\approx30(f/0.25)^{-1/2}\cm^{-3}$
%and the mean density is $\mn=f\nHc\approx7.5(f/0.25)^{1/2}\cm^{-3}$.
and the electron density $\nel$ in the cloudlets is $\sim36(f/0.25)^{-1/2}\cm^{-3}$.
The mass of the cloudlets inside the remnant amounts to $\sim110(f/0.25)^{1/2}\Msolar$.

From the volume EM of the hot component $(1-f)\nel\nHicm V\approx
1\E{58}\cm^{-3}$, the intercloud hydrogen number density obtained is
$\nHicm\approx0.9[(1-f)/0.75]^{-1/2}\cm^{-3}$.
About  $10[(1-f)/0.75]^{1/2}\Msolar$  of intercloud
gas is contained inside the remnant.

\subsection{Age and Explosion Energy}
Because the time since the gas was engulfed by the shock front is
contained in the ionization parameter $\nel t$, the value $\nel t\sim1.3\E{12}\cm^{-3}\s$ in the cool component implies an age of the
remnant: $t\gsim1.1\E{3}(f/0.25)^{1/2}\yr$, and $\nel
t\sim6.3\E{10}\cm^{-3}\s$ in the
hot component implies $t\gsim1.8\E{3}[(1-f)/0.75]^{1/2}\yr$, which are
both appreciably higher than the previous estimate of age, $600\yr$.
We consider that the blast wave propagates into the intercloud medium and
transmits a cloud shock into the cloudlets.  If the high temperature $T_{h}\sim2.6\keV$ (for the hot component in the NEI case) is
ascribed to the blast wave, the postshock temperature may be $\Ts=0.77T_{h}\sim2\keV$ (e.g.\ Rappaport, Doxsey, \& Solinger 1974).
The blast wave velocity would be
$\vs=(16k\Ts/3\mu\mH)^{1/2}\sim1.3\E{8}\cmps$ (where the mean atomic
weight $\mu=0.61$), and the age could be estimated as
$t=2r/5\vs\sim1.4\E{3}\yr$ using the Sedov model.  This estimate is
between the values given by the two ionization parameters.  Its
being smaller than one of the above values ($\gsim1.8\E{3}\yr$) is
understandable due to the severe departure from  spherically
symmetric evolution, the possible error in the determination of the
distance, and the uncertainty in the filling factor.  If the
temperature ($1.6\keV$) of the hot component obtained in the EI case
is used, the age estimate would be $\sim1.8\E{3}\yr$.  On the
assumption that the interaction with the cloudlets can be neglected in
the propagation of the blast wave, we estimate that the explosion
energy is
$E=(1.4\nHicm\mH/\xi)(r^{5}/t^{2})\sim2\E{50}[(1-f)/0.75]^{-1/2}\ergs$
(for $t\sim1.8\E{3}\yr$) where $\xi=2.026$.

\subsection{Issues Concerning the Inhomogeneous Medium}\label{sec:iss}

While the scenario of inhomogeneity reconciles the emission measures
of the two plasma components and produces estimates of the age
and explosion energy, it raises some questions to be discussed as follows.

The first is the pressure equilibrium between the ICM and the cloudlets
($\picm$ vs.\ $\pcl$) (McKee \& Cowie 1975).  From the values of
temperatures and densities obtained above, we have
$\picm/\pcl\sim(1/3)[3f/(1-f)]^{1/2}$.  Although here the filling
factor $f$ is adjustable, the ratio would be $1/3$ if $f\sim0.25$ is
taken as above.  It was recently suggested that intercloud
magnetic fields are a plausible pressure source for balancing the
cloudlets (Chevalier 1998).  Thus a mean magnetic field of
$\sim4.7\E{-4}\G$ would be needed inside the remnant, which is similar
to the observed strengths in other SNRs evolving in clouds (Claussen
et al.\ 1997).

The second is the thermal evaporation of the cloudlets within the
remnant.  Since the remnant is very likely to expand in a cloudy
medium, according to the model of White \& Long (1991), the X-ray
image should appear centrally brightened and smoothly darkened to the limb.
As will be shown in \S\ref{sec:mor}, the eastern concentration of 3C397 is
indeed near the center and hence one could not exclude the possibility of
evaporation. However its overall brightness distribution is much more
complicated than the model describes and the eastern concentration looks
compact and may be caused by other mechanism.  If the cloud evaporation is
unimportant here, a possible factor may be the inhibition of evaporation by the
electromagnetic instabilities (Levinson \& Eichler 1992) caused in the
collision of the ejecta with the cloudlets (Anderson \& Rudnick 1993).

The third is the photoevaporation of the cloudlets by the progenitor
star.  According to McKee, Van Buren, \& Lazareff (1984), a region of
radius $\sim28\parsec$ could be made homogeneous by the
photoionization of an O4-B0 star during its main sequence lifetime,
given an average medium density $\sim8\cm^{-3}$ as in the case of
3C~397.  Here, however, the cloudlets exist within a radius $2'$
($\sim4.7\parsec$). Hence the likelihood that the progenitor was a massive
early-type star is low and a type Ia supernova explosion
history may be favored.

The forth is the zero abundance of Si (with
an upper limit of about 0.1) in the hot component.  The hot component
is now ascribed to the intercloud gas behind the blast wave and
possibly contains most of the ejecta.  It has not yet reached
ionization equilibrium, and it is improbable that its Si is
completely depleted.  On the other hand, it is also difficult to
understand the absence of Si in the ejecta.  If the ejecta has a mass
of $\sim1\Msolar$ with a solar abundance of Si, its mass ratio
to the intercloud gas ($\sim0.1$) would entail the
absence of Si in the unshocked ICM.

\subsection{Ancient Records? The SGR Counterpart?}
The remnant age as young as $\sim1400$ --- $1800\yr$ is of interest in
connection with  ancient guest stars.  However, the extinction to
3C~397, $A_{\ssst V}\approx4.5\E{-22}\NH\sim13$mag (Gorenstein 1975) is
too high for ancient astronomers to have observed the SN explosion
with naked eyes, given the hydrogen column $\NH\sim2.9\E{22}\cm^{-2}$.
That may explain why there was no mention of a candidate historical record.
%(e.g. Clark \& Stephenson 1977).

Some researchers are interested in the seemingly compact central
source which could possibly harbor a neutron star (see Jones et al.\
1998).  3C~397 was also suggested as a possible candidate for the
extended \xray\ counterpart of SGR1900+14 (Greiner 1996).  However,
the absence of pulsed emission and power-law component in our analysis
seems not in support of these speculations.  It was recently proposed that a
quiescent \xray\ source, which may be associated with the SNR G42.8+0.6,
is responsible for the SGR (Hurley et al.\ 1999).

\subsection{Morphology}\label{sec:mor}

In the \ASCA\ SIS and \ROSAT\ HRI maps (Figure~2 and Figure~3), the
\xray\ emission is confined within the radio boundary, and again has
two concentrations similar to the \Einstein\ IPC map. (Here the maximum positional
error of the \ASCA\ images could be as large as about $40''$.) Two bright arcs in
the hard contour map are roughly coincident with the two
concentrations (see Figure~3).  Therefore the hot component is mainly
concentrated in the two bright portions.  The two hard arcs seem to
compose a bilateral structure in the western half.
The eastern half of the HRI image (Figure~3) seems to show a broken
bubble-like structure, with the eastern concentration of emission on
its western boundary.  The eastern \xray\ concentration which looks
like a compact source in the center of the remnant is coincident with
the intersecting portions of the east- and western bubble-like halves.
The hard contour map, the HRI image, and the
VLA map all show an outward protrusion on the eastern side. 

The elongation direction of the whole remnant is essentially
perpendicular to the galactic plane.  The high EM of the cool
component  suggests that the remnant is evolving in a cloudy
medium.  The brighter radio and \xray\ emissions on the west side
close to the galactic plane imply a density gradient of the
medium toward the plane, as has been speculated by Anderson \& Rudnick
(1993).  The $\NH$ values for  regions A and B obtained from the SIS
data and from the PSPC data using the VMEKAL code (\S\ref{sec:spec})
are also in favor of such a gradient. In fact the HII region G41.1$-$0.2
is known to be adjacent to the west (see \S\ref{sec:intro}).

The remnant morphology in the \xray\ images (especially the \ROSAT\
HRI image) seems to be suggestive of a bipolar (or a peanut-like)
bubble with roughly an east-west orientation, its symmetry axis not
perpendicular to the line of sight.
The western half may be a little more distant than the
eastern half, as implied by the $\NH$ values of regions A and B.
The western \xray\ concentration is located at the apex of the western
bright portion, which coincides with the western bright radio emission
rather well (see Figure~3).  This apex should result from the shock
wave interacting with the denser cloudy medium in the west.  On the other
side, the plasma may leak out of the the broken eastern bubble, so that
the X-ray emission from the eastern half is softer than from the
western half.  It is unclear why the remnant takes a bipolar shape.
The possibility that it was caused by two explosions seems low in
view of the similar sizes and similar plasma temperatures of the east
and west bubbles.  Another possibility may be that dense matter was
accumulated around the equatorial plane, but there is no direct
evidence for such a conjecture.  Two possible mechanisms could be
responsible for the matter accumulation.  One is a proto-stellar disk
of gas which was left over from the time of the formation of the
progenitor star (McCray \& Lin 1994).  The other could be the mass
loss of the progenitor star(s) along the equatorial plane; because the
progenitor was probably not a massive star, it could be a mass-losing binary
system and this is consistent with a type Ia supernova explosion
(that is mentioned in \S\ref{sec:iss}).

\section{Conclusion}
The \ASCA\ SIS0 and the \ROSAT\ PSPC spectral data on SNR 3C~397 are
analysed with a two-component NEI model using the SPEX code.  We also
fit simultaneously the \ASCA\ SIS0 and SIS1 spectra with the VMEKAL
model for an EI case.  The hard ($\gsim2\keV$) \xray\ emission is found to
arise primarily from the hot component which is responsible for S and
Fe~K$\alpha$ lines.  The cool component contributes dominantly
to the soft emission, responsible for Mg, Si, Fe~L lines. 
The cool component is found to
be approaching ionization equilibrium,  and its high emission measure
suggests that the remnant evolves in a cloudy medium.  The intercloud
magnetic fields may be a pressure source to balance the dense clouds.
The existence of an inhomogeneous surrounding medium implies that the
supernova progenitor may not be a massive early-type star. The zero
abundance of Si in the hot component needs further explanation.  We
did not find power-law, bremsstrahlung, or blackbody components in the
spectral analysis, nor pulsed signals in the temporal analysis.
We restored the \xray\ maps using the \ASCA\ SIS data and compared them
with the \ROSAT\ HRI and 20 cm VLA maps.  Two bright concentrations
might imply that the remnant with a bipolar structure encounters a
denser medium in the west.  The $\NH$ values obtained suggest a
distance of $\sim8\kpc$.  The Sedov model for the dynamics and the
ionization parameters for the two components yield an age
$\gsim1.4$--$1.8\E{3}\yr$, which is much greater than the previous
estimate of $600\yr$.

\acknowledgements{We would like to thank J.\ Tr\"{u}mper and R.\ McCray
for critical comments during the preparation of this
manuscript. We would also like to thank J.\ Condon for critical reading and
comments of the manuscript, and J.-H.\
Huang, Q.-S.\ Gu, and W.\ Becker for technical help.  Data analysis is
carried out on the SUN workstation at the Laboratory for Astronomical
Data Analysis of the Department of Astronomy, Nanjing University.
This work is supported by a grant from the NSF of China, a grant from
the Ascent Project of the State Scientific Ministry of China, and a
grant from the State Education Ministry of China for scholars
coming back from abroad.}

\clearpage

\begin{center}
\section{Figure captions}
\end{center}

\figcaption[fig1.ps]{The \xray\ spectrum extracted from the SIS0 data,
fitted with the two component NEI model in the SPEX
code. The dotted line represents the contribution of the cool component
in the model, the dashed line represents the contribution of the hot
component, and the solid line is the sum of the both.
}

\figcaption[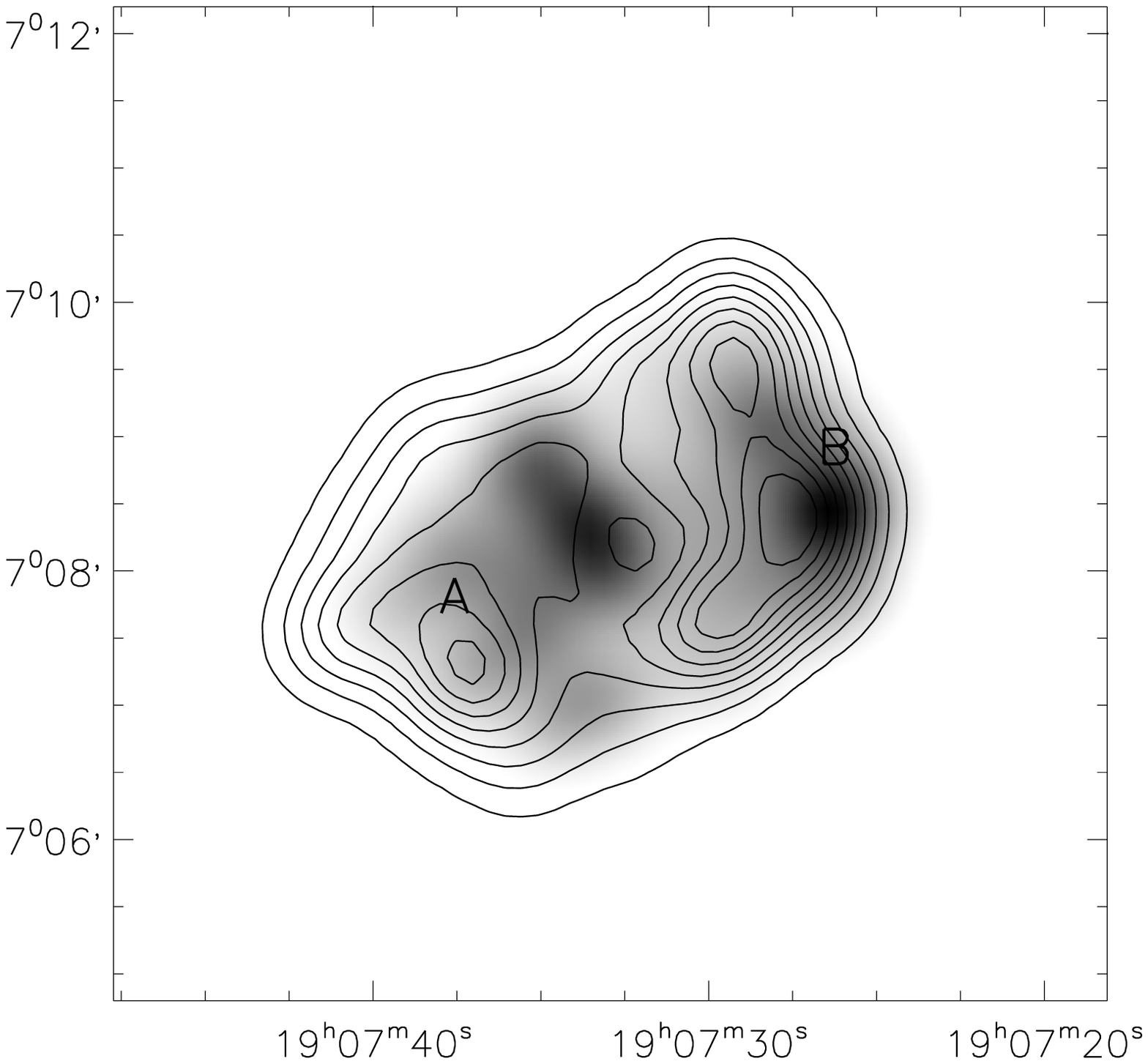]{Smoothed \xray\ (0.5 --- 10 keV) image of 3C~397 restored from the
\ASCA\ SIS data, with an overlay of the 20 cm radio emission contours.
The contours are from 0.1 to 0.9 of maximum  with 8 linear intervals.}

\figcaption[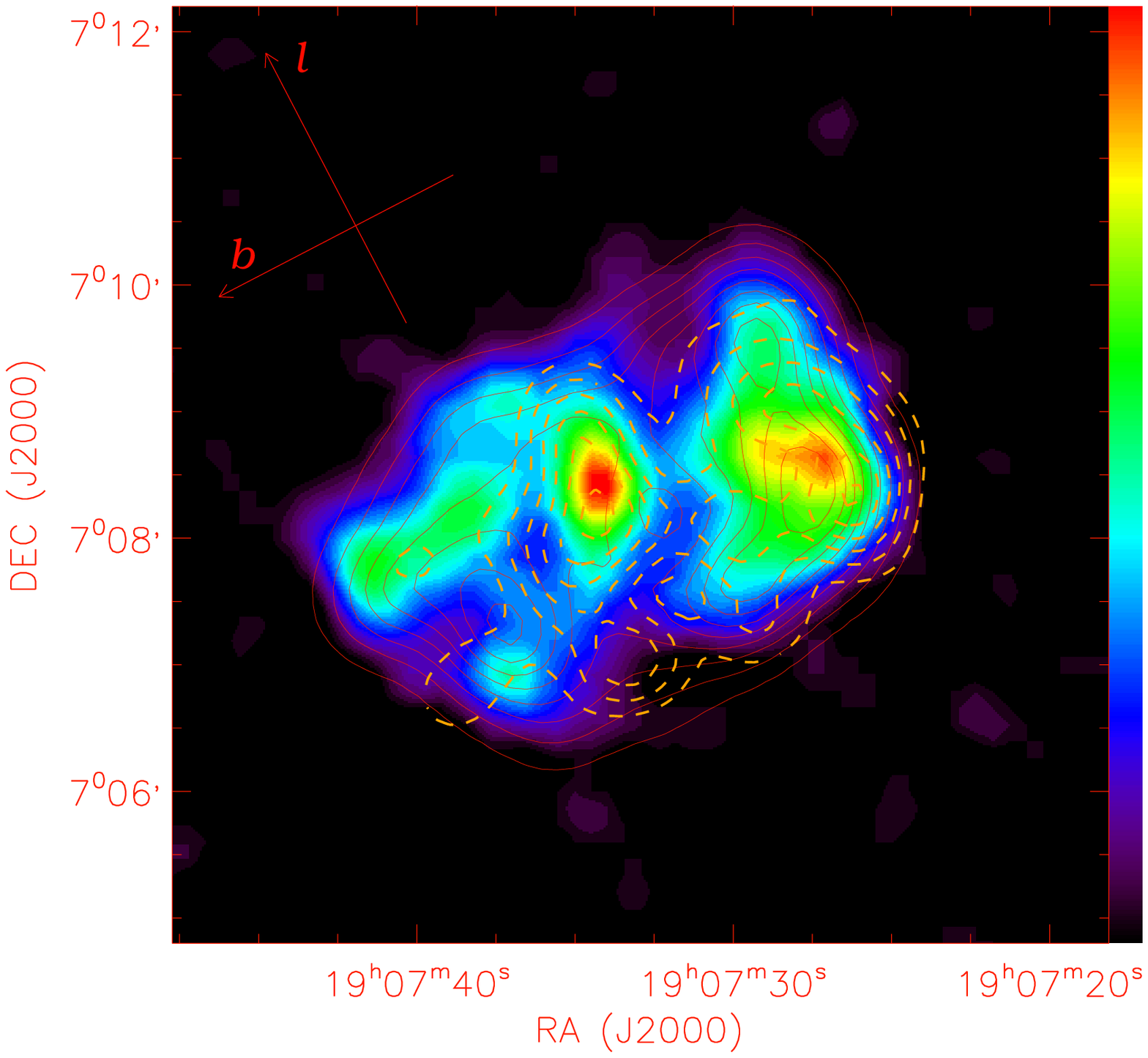]{Smoothed \ROSAT\ HRI image superposed with the
restored SIS 2---10 keV (hard) contours (dashed lines) that are from 0.15 to
0.9 of maximum with 5 linear intervals and the 20 cm radio emission
contours (solid lines) that are from 0.1 to 0.9 of maximum with
8 linear intervals.}

\clearpage

\centerline{\begin{tabular}{c|ll|l}
\multicolumn{4}{c}{TABLE 1}\\
\multicolumn{4}{c}{Fitting results obtained with the NEI model by applying SPEX to}\\
\multicolumn{4}{c}{the \ASCA\ SIS0 and \ROSAT\ PSPC spectral data. The element abundances}\\
\multicolumn{4}{c}{for the two temperature components are independent and coupled, respectively.}\\
\multicolumn{4}{c}{The 90\% confidence ranges ($\Delta\chi^{2}=2.706$) are indicated.}\\ \hline\hline
      & \multicolumn{2}{c|}{independent case} & \multicolumn{1}{c}{coupled case} \\ \cline{2-4}
      & \multicolumn{1}{c}{SIS0} & \multicolumn{1}{c|}{SIS0+PSPC} & \multicolumn{1}{c}{SIS0} \\ \hline
\multicolumn{1}{c}{\bf hot component}\\
$n_{e}n_{H}V$ ($10^{58}\cm^{-3}$) & $1.29^{+0.33}_{-0.18}$ & $1.33^{+0.40}_{-0.14}$ & $0.99^{+1.06}_{-0.16}$ \\
$T$ (keV) & $2.63^{+0.43}_{-0.53}$ & $2.53^{+0.31}_{-0.57}$ & $3.84^{+0.00}_{-0.04}$ \\
$n_{e}t$ ($10^{12}\cm^{-3}\s$) & $6.26^{+3.83}_{-1.63}\E{-2}$ & $6.49^{+4.49}_{-1.61}\E{-2}$ & $5.01^{+0.14}_{-0.36}\E{-2}$ \\
{[Mg/H]} & $1.89^{+1.36}_{-1.89}$ & $1.74^{+1.49}_{-1.53}$ & $0.69^{+0.00}_{-0.00}$ \\
{[Si/H]} & $0.00^{+0.08}_{-0.00}$ & $0.00^{+0.19}_{-0.00}$ & $0.67^{+0.05}_{-0.00}$ \\
{[S/H]} & $1.88^{+0.49}_{-0.64}$ & $1.87^{+0.41}_{-0.70}$ & $1.98^{+0.02}_{-0.00}$ \\
{[Fe/H]} & $2.45^{+1.74}_{-0.67}$ & $2.65^{+2.28}_{-0.62}$ & $0.91^{+0.00}_{-0.00}$ \\
\multicolumn{1}{c}{\bf cool component}\\
$n_{e}n_{H}V$ ($10^{58}\cm^{-3}$) & $344^{+232}_{-88}$ & $ 341^{+298}_{-79}$ & $490^{+489}_{-29}$\\
$T$ (keV) & $0.239^{+0.010}_{-0.015}$ & $0.238^{+0.007}_{-0.019}$ & $0.226^{+0.003}_{-0.053}$\\
$n_{e}t$ ($10^{12}\cm^{-3}\s$) & $1.34^{+\infty}_{-0.98}$ & $1.38^{+\infty}_{-1.09}$ & $2.67^{+0.07}_{-2.11}$ \\
{[Mg/H]} & $0.38^{+0.18}_{-0.16}$ & $0.37^{+0.16}_{-0.19}$ & ------ \\
{[Si/H]} & $0.94^{+0.35}_{-0.25}$ & $0.91^{+0.33}_{-0.10}$ & ------ \\
{[S/H]} & $0.82^{+2.48}_{-0.82}$ & $0.83^{+1.82}_{-0.83}$ & ------ \\
{[Fe/H]} & $0.33^{+0.41}_{-0.28}$ & $0.34^{+0.54}_{-0.31}$ & ------ \\ \hline
$N_{H}$ ($10^{22}\cm^{-2}$) & $2.87^{+0.12}_{-0.12}$ & $2.87^{+0.20}_{-0.13}$ & $3.04^{+0.00}_{-0.05}$ \\
$\chi^{2}/{\rm d.o.f.}$ & $341/254$ & $579/459$ & $416/258$ \\ \hline
\end{tabular}}
\bigskip

\centerline{\begin{tabular}{c|lll}
\multicolumn{4}{c}{TABLE 2}\\
\multicolumn{4}{c}{Fitting results obtained with the EI model by applying XSPEC}\\
\multicolumn{4}{c}{simultaneously to the SIS0 and SIS1 data. The 90\% confidence ranges}\\
\multicolumn{4}{c}{($\Delta\chi^{2}=2.706$) are indicated.}\\ \hline\hline
      & whole & portion A & portion B \\ \hline
\multicolumn{1}{c}{\bf hot component}\\
$n_{e}n_{H}V/(d/8\kpc)^{2}$ ($10^{58}\cm^{-3}$) & $2.04^{+0.26}_{-0.27}$ & $0.59^{+0.07}_{-0.14}$ & $0.99^{+0.11}_{-0.19}$ \\
$T$ (keV) & $1.67^{+0.11}_{-0.09}$ & $1.55^{+0.19}_{-0.14}$ & $1.23^{+0.08}_{-0.06}$ \\ 
{[Mg/H]} & $7.67^{+4.26}_{-2.97}$ & $8.19^{+8.24}_{-4.32}$ & $0.00^{+2.51}_{-0.00}$ \\ 
{[Si/H]} & $0.00^{+0.10}_{-0.00}$ & $0.00^{+0.18}_{-0.00}$ & $0.00^{+0.33}_{-0.00}$ \\ 
{[S/H]} & $0.60^{+0.32}_{-0.29}$ & $0.52^{+0.62}_{-0.51}$ & $1.00^{+0.78}_{-0.46}$ \\ 
{[Fe/H]} & $3.72^{+0.65}_{-0.56}$ & $3.73^{+1.94}_{-1.06}$ & $5.57^{+3.02}_{-1.37}$ \\ 
\multicolumn{1}{c}{\bf cool component}\\
$n_{e}n_{H}V/(d/8\kpc)^{2}$ ($10^{58}\cm^{-3}$) & $521^{+76}_{-61}$ & $117^{+32}_{-26}$ & $197^{+60}_{-38}$\\
$T$ (keV) & $0.223^{+0.004}_{-0.003}$ & $0.224^{+0.008}_{-0.007}$ & $0.222^{+0.006}_{-0.033}$\\ 
{[Mg/H]} & $0.58^{+0.08}_{-0.06}$ & $0.61^{+0.14}_{-0.13}$ & $0.48^{+0.15}_{-0.11}$\\ 
{[Si/H]} & $0.97^{+0.12}_{-0.11}$ & $1.10^{+0.27}_{-0.23}$ & $0.97^{+0.20}_{-0.17}$\\ 
{[S/H]} & $5.60^{+1.03}_{-0.83}$ & $6.48^{+2.63}_{-1.83}$ & $4.12^{+1.79}_{-1.60}$\\ 
{[Fe/H]} & $0.00^{+0.15}_{-0.00}$ & $0.00^{+0.25}_{-0.00}$ & $0.00^{+0.35}_{-0.00}$\\ \hline
$N_{H}$ ($10^{22}\cm^{-2}$) & $2.82^{+0.07}_{-0.04}$ & $2.66^{+0.12}_{-0.07}$ & $3.08^{+0.27}_{-0.06}$\\ 
$\chi^{2}/{\rm d.o.f.}$ & $1142/559$ & $613/559$ & $593/559$\\ \hline
\end{tabular}}

\end{document}